# Design and Simulation of Si-Photonic Nanowire-Waveguides with DEP Concentration Electrodes for Biosensing Applications

[1]Anders Henriksson*, [1]Peter Neubauer, [2] Mario Birkholz

[1] Chair of Bioprocess Engineering, Department of Biotechnology, Technische Universität Berlin, Ackerstraße 76, 13355 Berlin, Germany;

[2] IHP—Leibniz-Institut für Innovative Mikroelektronik, Im Technologiepark 25, 15236 Frankfurt (Oder), Germany

* Correspondence: henriksson@campus.tu-berlin.de

**Abstract**

Silicon-based photonic biosensors, such as microring resonators and Mach-Zehnder interferometers, offer significant potential for the detection of analytes at low concentrations. To enhance response time and improve the limit of detection within practical time scales, dielectrophoresis (DEP) has been proposed as a viable solution. In this approach, two electrodes are placed in close proximity to the sensor surface. By applying an AC field, analytes are transported to the sensor surface due to the polarization of the solvent and the particles, effectively overcoming the diffusion barrier.

In this communication, we explore various possibilities for realizing DEP electrodes for nanowire waveguides using commercially available photonic integrated circuit (PIC) technology. Finite Element Method (FEM) simulations suggest that the most beneficial electrode configuration is a planar electrode geometry on the device layer combined with a second electrode pair on the metal 1 layer.

## 1.Introduction

Silicon-based photonic biosensors, such as microring resonators and Mach-Zehnder interferometers, offer significant potential for detecting analytes at low concentrations.[1,2] Their advantages include label-free sensing, compatibility with state-of-the-art semiconductor technology, and straightforward strategies for mass production and miniaturization. However, like all receptor-based biosensors, effective detection requires that analytes be transported to the sensor surface and captured by the receptors. This step relies on diffusion, which can limit response time and detection limits within a reasonable time scale.[3]

In the past decade, several studies have explored the use of dielectrophoresis (DEP) to transport analytes to the sensor surface, showing promising results that could increase sensitivity by several orders of magnitude and significantly reduce response time.[4] Dielectrophoresis involves applying an inhomogeneous AC field that polarizes both the solvent and particles, causing a net movement of the particles either toward the strongest electric field gradient or away from it, depending on the applied frequency and the dielectric properties of the particles. By placing two electrodes in close proximity to the sensor surface, it is possible to focus analytes onto the sensor surface, thereby enhancing the system's sensitivity.

To implement dielectrophoresis in silicon photonics, two main electrode configurations can be realized using commercially available photonic integrated circuit (PIC) technology.[5] One approach is to place the electrodes on the metal 1 layer, which is approximately 1.6 µm above the sensor surface. Another approach is to place the electrodes on the device layer using doped silicon structures.

In this communication, we explore different geometries for realizing DEP electrodes on silicon nanowire waveguides, aiming to improve the sensitivity of microring resonator biosensors.

## 2. Method

FEM Simulations:

Finite Element Method (FEM) simulations were performed using COMSOL Multiphysics with the AC/DC module. The material properties utilized in the simulations are detailed in Table 1. This setup allowed us to model the electric field distribution and the resulting dielectrophoretic forces acting on the analytes, providing insights into the optimal electrode configurations for enhancing biosensor sensitivity. The materials are taken from several databases[6] with the relative permittivity of PBS puffer set to resemblance water with a value of 80.[7,8] The applied AC voltage was set to 10 Vpp in the simulations.

Table 1. Material properties used in finite element simulations.

| Material | Relative Permittivity | Electric Conductivity (S/m) |
| --- | --- | --- |
| $SiO_2$ | 3.74 | $10^{-14}$ |
| Medium (PBS Puffer) | 80 | 1 |
| Si | 11.7 | 10 |
| $n^+$-Si | 11.7 | $2 \times 10^4$ |
| Metal 1 (aluminum) | 1 | $10^5$ |

# 3. Results

To enhance the sensitivity of nanowire waveguides in silicon photonic integrated circuits (PICs) using dielectrophoretic concentration electrodes, there are two main approaches compatible with commercially available technology. One method involves placing electrodes on the device layer in close proximity to the core sensor waveguide, as illustrated in Figure 1a. In this setup, the strongest $\nabla E^2$ gradient is observed at the edges of the electrodes. However, a notable local maximum can also be observed on the core sensor waveguide between the electrodes. This phenomenon resembles the mechanism of insulating dielectrophoresis (iDEP), where structures like posts, membranes, obstacles, or constrictions within microfluidic channels create localized electric field gradients. These structures distort the applied electric field, creating a high electric field gradient with a local maximum within the channel. In the electrode structure described, the core waveguide functions as an insulating obstacle that generates this local maximum.

The spacing between electrodes critically influences the strength and distribution of the electric field. When electrodes are positioned within 0.5 μm of the core waveguide, the electric field gradient is evenly distributed around it. However, increasing the distance between electrodes concentrates the electric field gradient at the edges while weakening it near the core waveguide. Conversely, placing electrodes too close to the waveguide risks evanescent coupling, leading to significant optical loss.[9] Therefore, optimizing the electrode distance is essential to maximize $\nabla E^2$ at the core sensor waveguide while minimizing optical losses.

Placing the electrodes on the metal 1 layer is another effective method for implementing DEP electrodes in a PIC.[10] These electrodes offer greater design flexibility, and as shown in simulation ii in Figure 1c, the $\nabla E^2$ extends into the bulk solution, enabling the capture of analytes from a greater distance. However, the strongest $\nabla E^2$ in this configuration is relatively far from the core waveguide, meaning analytes are transported to the electrodes rather than directly to the sensor waveguide. The primary purpose of these electrodes is to capture analytes from further away and concentrate them closer to the sensor. Once preconcentrated, these analytes can interact more effectively with the device layer electrodes, which are designed to transport particles directly onto the core waveguide for detection.

Simulation iii in Figure 1c demonstrates an approach that combines the two previous configurations. In this setup, electrodes on the device layer and the metal 1 layer are used simultaneously. This configuration has the advantage of generating the strongest $\nabla E^2$ at the core sensor waveguide due to the device layer electrodes, while also allowing the $\nabla E^2$ to penetrate far into the bulk solution thanks to the metal 1 layer electrodes. This combined approach thus has a great potential to enhance the performance of the sensor taking constructive advantage of both previously approaches.

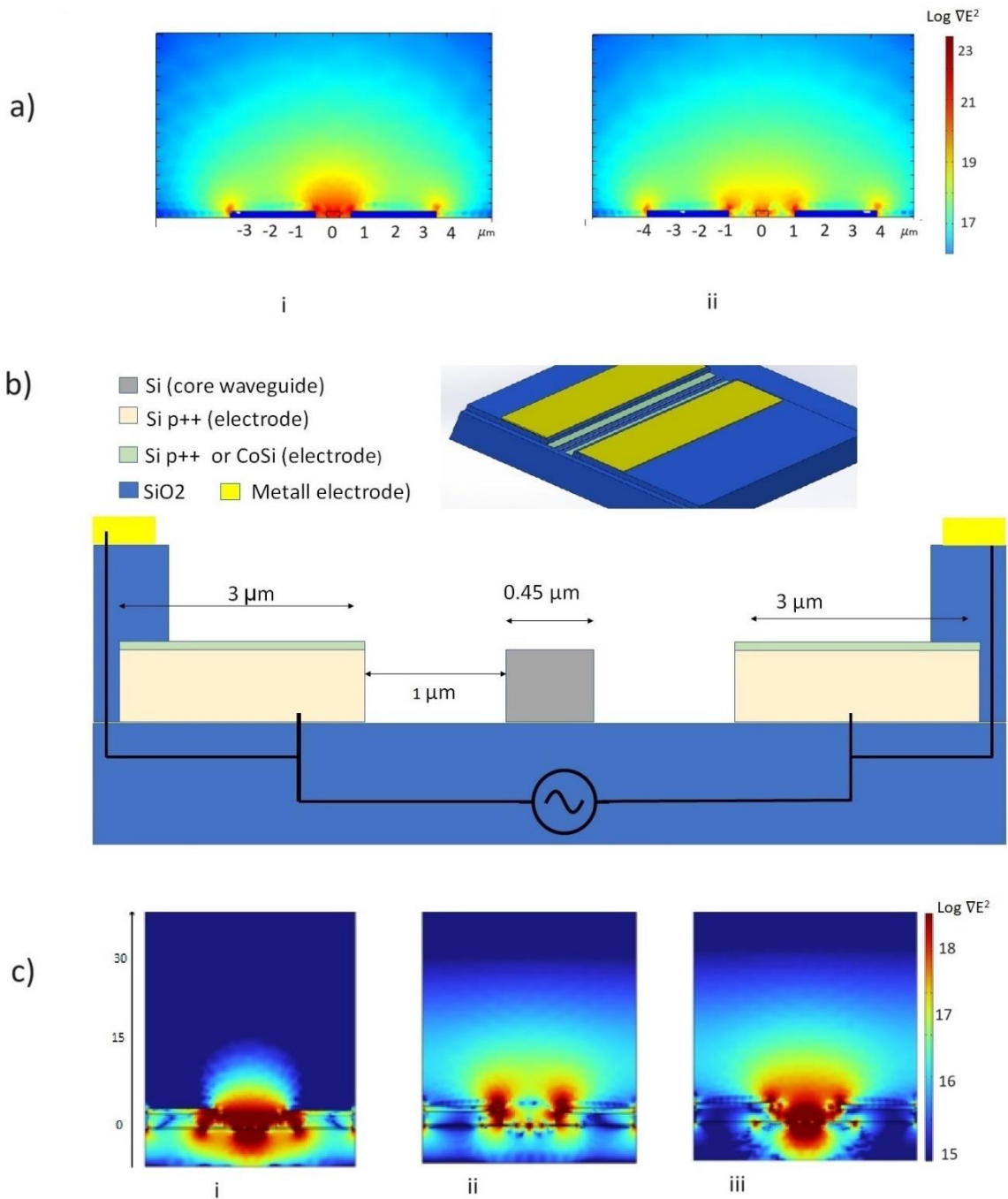

Figure 1. A proposed electrode geometry with a coplanar electrode pair on the device layer next to the sensing waveguide. a) simulated $\nabla E^2$ in (log($\nabla E^2$ /V²m⁻³)) with the electrodes placed at a distance of i) 0.5 μm and ii) 1 μm from the waveguide. b) cross- section and top view of an electrode pair solution, one placed at the device layer and one at the metal 1-layer. c) Simulated $\nabla E^2$ with i) only the electrodes on the device layer turned on. ii) only the electrodes on the metal 1-layer. iii) electrodes on the device layer and metal 1-layer simultaneously turned on.

## 4. Conclusion

In our study, we investigated integrating DEP electrodes into silicon photonic integrated circuits (PICs) to enhance nanowire waveguide sensitivity for biosensing. Through FEM simulations using COMSOL Multiphysics, we evaluated two electrode configurations: on the device layer near the core sensor waveguide and on the metal 1 layer.

Electrodes on the device layer create a strong, uniform electric field gradient around the waveguide, with a local maximum similar to insulating dielectrophoresis (iDEP). However, careful electrode spacing is crucial to avoid evanescent coupling.

Metal 1 layer electrodes offer flexibility and extend the electric field into the bulk solution, capturing analytes from a distance. While the strongest field gradient is farther from the waveguide, these electrodes may potentially concentrate analytes near the sensor for efficient detection.

Combining both configurations optimizes performance by generating a strong electric field gradient at the core sensor waveguide and extending it into the bulk solution. This hybrid approach may enhance analyte capture and transport efficiency.

The findings presented in this communication highlight the potential of integrating DEP electrodes into PICs to advance biosensing capabilities, with further validation needed for practical application optimization.

## 5. References.

**Acknowledgement**

This research was funded by the German Research Foundation DFG via the grant HE 8042/1-1.